# GC-MS Profile of *Diodella sarmentosa (SW) Bacigalupo El Cabral ex Borhidi* Ethanol Leaf Extract and its Total Dehydrogenase Inhibitory Potential

*Callistus Izunna Iheme[1], Doris Ifeyinwa Ukairo[2], Okechukwu Igwe Oguoma[3], Linus Ahumareze Nwaogu[2], Amanda Ugochi Ezirim[2], Chiamaka Perpetua Nzebude[2] and Chidimma Caryn Ezerioha[2]

[1]*African Center on Future Energies and electrochemical system (ACE-FUELS), Federal University of Technology Owerri, Imo State, Nigeria.*

[2]*Department of Biochemistry, Federal University of Technology Owerri, Imo State, Nigeria*

[3]*Department of Microbiology, Federal University of Technology Owerri, Imo State, Nigeria*

*Corresponding author's Email: callistus.iheme@futo.edu.ng ; Tel: +234 (0)7031014133.

**Abstract**

Phytochemical composition of ethanol leaf extract of *Diodella sarmentosa* was profiled with GC-MS and the inhibitory property of the extract against total microbial dehydrogenases were assessed. The major constituents of the extract were squalene (29.50%), Phytol (24.68%), phenol, 3-pentadecyl- (18.58%), 1-Butanol, 3-methyl- (9.09%) and n-Hexadecanoic acid (7.78%). The minimum inhibitory concentration (MIC) of the extract against broad spectrum of microbial population was assessed. *Bacillus subtilis, Candidas spp*, and *Penicillium spp* were more sensitive to the treatment and thus; were further investigated using Dehydrogenase activity assay method. Total dehydrogenase activities of *Bacillus subtilis, Candidas spp*, and *Penicillium spp* at the extract concentration range of 0 to 2000mg/ml were progressively inhibited at increasing extract concentrations. The threshold inhibitory concentrations ($IC_{50}$) of the extracts against *Candidas spp, Penicillium spp* and *Bacillus subtilis* were 275μg/ml, 322μg/ml and 411μg/ml respectively. Our findings suggested the extract as a useful source of antimicrobial phytochemicals for pharmaceutical use.

## 1.0 Introduction

Increasing cases of resistance and side effects have been reported of synthetic antimicrobial drugs. Consequently, attentions are now being shifted towards herbal sources for alternative medicinal products. However, it has been reported that the viability of microbial population depends on oxido-reduction metabolism [Salazar *et al*., 2011] which is in turn catalyzed by oxido-reductases (eg dehydrogenase EC1.1.1). Therefore, the assessment of the total dehydrogenase inhibitory property of the extract is a step towards the identification of the bio-active compounds in *Diodella sarmentosa (SW) Bacigalupo El Cabral ex Borhidi* leaf extract since the leaf is used traditionally among the Igbos for the treatment of topical infections. This is the first time the bio-active phytochemical compositions of the plant have been profiled.

## 2.0 Results and Discussion

Table 3.1 showed the GC-MS profile of phytochemical compositions of *Diodella sarmentosa (SW) Bacigalupo El Cabral ex Borhidi* ethanol leaf extract. The result indicated that the ethanol leaf extract were rich in such bioactive compounds like squalene (29.50%), Phytol (24.68%), phenol, 3-pentadecyl- (18.58%), 1-Butanol, 3-methyl- (9.09%) and n-Hexanoic acid (7.78%). Such other compounds like oxetane,2,4-dimethyl-, trans- (1.63%), Decane, 4,5-dibromo-($R^*$,R*)- (2.94%), androstan-3-ol, 9-methyl-, acetate, (3β, 5α) (3.17%), and 4-(-4-) (2-hydroxybenzoyl) amino (anilino)-4-oxo (1.67%) were found in relatively small quantity. Other compounds found in trace quantities were 7-Azabicyclo(-4.1.0) heptane (0.56%) and methyl 2,4-dimethyltetradecanoate (0.56%). The observed antimicrobial activity of the extract may have been caused by n-hexadecanoic acid, phytol and 1-Butanol, 3-methyl (Table 3.2). The highest dehydrogenase activity was recorded for *Penicillium spp,* followed by *Candidas spp* and the least being *Bacillus substilis (*Table 3.3). Variations in activities among the microbial strains maybe connected to the physiological variations of the dehydrogenase system among microbial populations (Praven-Kumar, 2003). At the Threshold inhibitory concentrations of 275μg/ml, 322μg/ml and 411μg/ml, 50% activities of the enzyme from *Candidas spp, Penicillium spp* and *Bacillus subtilis* respectively were inhibited (Table 3.4). This suggested that the extract had more effect on fungi than on the bacterium strain; and n-Hexadecanoic acid which has been reported to have reductase inhibitory potential may be responsible for this observed effect (Table 3.2). At 100mg/ml, significant inhibitory zones were observed against *Bacillus subtilis, Candidas spp,* and *Penicillium spp* (Fig 3.1). The inhibitory effect of the extract on the organisms may have been accounted for by the presence of Phytol, n-Hexadecanoic acid, and 1-Butanol,3-methyl- which have been reported to have antimicrobial properties (Table 3.2). The effect of the extract on the different strain of the organisms were concentration-dependent and differed markedly among the studied organisms (Figure 3.2). The activities of the enzyme strongly correlated with the extract concentrations (Fig 3.3). Figure 3.3A, Figure 3.3B and Figure 3.3C were Linearized plots of Log DHA Activities vs Extract Concentration (μg/ml) for A, B and C respectively, where A= *B.subtilis*, B= *Candidas spp*, and C= *Penicillum spp*. The higher value $R^2$ 0.981 ($0.948 < R^2 \leq 0.981$) confirmed that the concentration of the extract is inversely related to the activities of the enzyme (Fig 3.3). The implication of which is, at increased extract concentrations, the respiratory and carbon metabolism of the organisms would be negatively affected. This finding corroborated with those of [Osadebe and Ukueze, 2004]. From our findings, we concluded that ethanol extract of *Diodella sarmentosa (SW) Bacigalupo El Cabral ex Borhidi* can be a useful source of pharmaceutical and cosmetic bioactive compounds. However, further research will be required to determine the mechanism of the bio-active compounds action against the microbial isolates.

### 3.0 Experimental

**3.1 Sample Collection and Preparation:** The leaves of *Diodella sarmentosa (Sw) Bacigalupo El Cabral ex Borhidi* were collected from the Federal University of Technology Owerri, Imo State, in the month of April 2016, and the plant was identified by [Francis, 2016]. A voucher number of the plant was 001/FWT/FUTO/2016 and was deposited in the Herbarium. The fresh leaves were thoroughly washed with clean running tap water 2 – 3 times and then air-dried at room temperature for two weeks. The dried samples were crushed and preserved at room temperature in an air-tight container. Twenty grams (20g) of the sample was mixed with 500ml of ethanol at the temperature range of 60 to 65$^0$C for 24 hrs using soxhlet extractor. Using rotary vacuum evaporator, the solvent was evaporated to 23% yield. The resulting viscous semisolid extract was stored in Petri dishes wrapped with aluminum foils in a refrigerator at -4$^o$C for further analysis.

**3.2 GC-MS ANALYSIS:** The samples were subjected to GC-MS analysis on GC-MS equipment (Thermo Scientific Co.) Thermo GC-TRACE Ultra Ver. 5.0, Thermo MS DSQ II. Experimental conditions of GC-MS system were adhered; DBS-MS Capillary Standard Non-polar Column, dimension: 30Mts, ID: 0.25mm, Film thickness: 0.25μm. Flow rate of mobile phase (carrier gas: He) was set at 1.0ml/min. In the gas chromatography part, temperature program (oven temperature) was 70$^0$C raised to 260$^0$C at 6$^0$C/min and injection volume was 1μl. Samples dissolved in ethanol were run to completion at a range of 50-650m/z and the results were compared to Wiley Spectral Library Search Program. Interpretation of the mass spectrum of GC-MS was carried out with the database of National Institute of Standard and Technology (NIST). The mass spectrum of the known components in the NIST Library were compared to the unknown spectrum from the extract. The relative % amount of each component was calculated by comparing the average peak area to the total area. The GC-MS analysis chromatogram was presented in Fig 3.1, and the list of constituents in Table 3.1. The table 3.2 presented the major components and their biological activities.

**3.3 Assessment of Anti-microbial Characteristics of the Plant Extract**

The modified agar well diffusion method of Collins *et al.*, [1995] was applied to estimate the anti-microbial properties of the ethanol leaf extract. Different concentrations of the extract, 100mg/ml, 50mg/ml, 25mg/ml,12.5mg/ml and 6.25mg/ml were prepared. Subsequently, 0.1ml of the standard 24hours old culture of the test organisms in nutrient and potato dextrose broths were spread unto sterile nutrient, and potato dextrose agar plates (Muller – Hintine Agar) while allowing them to set [Junaid, 2006]. In addition, wells of approximately 5mm in terms of diameter on the plates were bored with the help of a sterile cork borer. The concentration of 0.05ml each of the extract was dispensed down the wells and then allowed to stay for 15minutes for the diffusion of the extract. These were then incubated at 37ºC for 24 h. Microorganisms that showed high significant zone of inhibition level were subjected to dehydrogenase activity assay using the modified method of Nweke *et al.,*[2007].

### 3.4 Total Microbial Dehydrogenase Inhibitory Evaluation and Statistical Analysis

The activities of total dehydrogenases were assessed using the method of Nweke *et al.,* [2007] as modified by Alisi *et al.*, [2008]. Accordingly, 2,3,5 – triphenyl tetrazolium chloride (TTC) was used as the artificial electron acceptor, which on acceptance of electrons, was made red in coloured triphenylformazan (TPF). The work was done in 4ml volumes of nutrient broth – glucose-TTC-medium and dextrose broth-glucose-TTC medium for the bacterial and the fungal strains, respectively. The media were aided with different concentrations (0-2000mg/ml) of the leaf extract using separate screw-capped test (experimental) tubes. Final extract concentrations of 0, 50, 100, 200, 400, 800, 1600, and 2000 μg/ml were obtained by adding 1ml of 0.4% (w/v) TTC in deionized water to each tube. Isolates and the media devoid of *Diodella sarmentosa (Sw) Bacigalupo El Cabral ex Borhidi* extracts served as the controls. At room temperature (28 ± 2ºC) for 16h, the reaction mixture was further incubated. The triphenyl formazan exhibited was obtained in 4ml of amyl alcohol at 500 nm spectrophotometrically. By using a dose-response curve (0-20mg/ml TPF (sigma) in amyl alcohol), the amount of formazan yielded was computed. Linear plots of the percentage inhibition for each test organism against the concentration of the extracts using gamma parameters were plotted at various inhibition rates based on linear regression.

The values gotten in triplicates were placed on a two-way ANOVA and were expressed as Mean ± standard deviation. The values at P<0.05 were considered statistically significant. Linear regression models were applied for the determination of the $IC_{50}$ of the extracts on the organisms.

### 4.0 Conclusion

***Diodella sarmentosa (SW)*** has been used over several years among Ndi-Igbo from South eastern Nigeria for the treatment of topical microbial infection without scientifically backed evidence on its efficacy. Our study, being the first to profile the leaf extract, has provided a substantive scientific evidence that the ethanol extract of the plant is rich in such antimicrobial compounds as phytol, n-Hexadecanoic acid, and 1- Butanol, 3-methyl. The extract was also found to contain high concentration of squalene, a compound with emollient, antioxidant, moisturizing and antitumor property. From our findings, we concluded that ethanol extract of ***Diodella sarmentosa (SW)*** can be a useful source of pharmaceutical and cosmetic bioactive compounds. However, further research will be required to determine the mechanism of the bioactive compounds action against the microbial isolates.

**Acknowledgement:** The authors wish to acknowledge the technologists in the Department of Biochemistry and Microbiology, Federal University of Technology, Owerri, for their technical supports.

**Declaration of Interest:** There is no conflict of interests among the authors.

### References

Alisi CS, Nwanyanwu CE, Akujobi C O and Ibegbulem C O (2008). Inhibition of dehydrogenase activity in pathogenic bacteria isolates by aqueous extract of Musa paradisiacal (varsapientum). *Afri J. Biotechnol*;**7**(12): 1821-1825.

Ammal RM and Bai GV. (2013). GC-MS Determination of bioactive constituents of Heliotropium indicum leaf. *Journal of Medicinal Plants*, **1**(6), 30-33.

Bawankar R, Deepti VC, Singh P, Subashkumar R, Vivekanandhan G and Babu S. (2013). Evaluation of bioactive potential of an Aloe vera sterol extract. *Phytotherapy Research*, **27**(6): 864-868.

Collins CH, LynesP M and Grange J M (1995). Microbiological method.Butterwort-Heinemann Ltd. Britain. *Microbiology Method (7th edition)*; **33:**. 175 – 190.

Correia SJ, David JP, and David JM (2006). Secondary metabolites of species of Anacardiaceae. Quim. *Nova* **29**(6):1287-1300.

Farina M., Preeti B and Neelam P (2014). Phytochemical Evaluation, Antimicrobial Activity, and Determination of Bioactive Components from Leaves of *Aegle marmelos*, *BioMed Research International*, **2014**: 1-11.

Junaid SA, Olabode AO, Onwuliri FC, Okwori AE J and Agina SE (2006). The antimicrobial properties of *Ocimumgratissimum* gastrointestinal isolates. *Afri. J. Biotechnol*; **22**:2315-2321

Kalvodova L. (2010). Squalene-based oil-in-water emulsion adjuvants perturb metabolism of neutral lipids and enhance lipid droplet formation. Biochemical and biophysical research communications, **393**(3): 350-355.

Khasawneh MA, Elwy HM, Hamza AA, Fawzi NM, and Hassan AH. (2011). Antioxidant, anti-lipoxygenase and cytotoxic activity of Leptadenia pyrotechnica (Forssk.) decne polyphenolic constituents. *Molecules*, **16**(9): 7510-21.

Nweke CO, Alisi CS, Okolo JC and Nwanyanwu CE (2007). Toxicity of zinc to heterotrophic bacteria from tropical river sediment. *Appl Ecol. Environ Research*; **(1):** 123 – 132.

Osadebe PO and Ukueze SE (2004). A comparative study of the phytochemical and antimicrobial properties of the Eastern Nigeria Species of African Mistletoe. *J.Biol.Res. Biotechnol*; **2**(1):18-23.

Praveen – Kumar JC (2003). 2,3,5-triphenyl tetrazolium chloride (TTC) and electron acceptor of culturable cell bacteria, fungi and antinomycetes Boil. *Fert.Soil*; **28**:186 -189.

Salazar S, Sanchez L, Alvarez J, Valverde A, Galindo P, Igual J, Peix A and Santa-Regina I (2011). Correlation Among Soil Enzyme Activities Under Different Forest System Management Practices. *Ecological Engineering*, **37**: 1123-1131.

**Figure Captions**

Figure 3.1: Diameter of Zone of Inhibition (mm) : Maximum zone of inhibition was observed at 100mg/ml of the extract.

Fig. 3.2: DHG activity of the organisms vs extract concentration (μg/ml): The enzyme activity was dose dependent.

Figure 3.3A, B&C**:** Linearized plot of Log DHA Activities vs Extract Concentration (μg/ml)

The $R^2$ values of 0.9483, 0.9544, and 0.9819 for *B.subtilis, Candidas spp, and Penicillum spp* respectively, is indicative of the strong correlation of the plant extract against the enzyme.

The positive values further elucidate the inverse relationship between the extract and the enzyme.

## 3.0 RESULTS

Table S3.1 GC- MS Analysis of Phytochemical Composition of Ethanol Extract of *Diodella sarmentosa (Sw) Bacigalupo El Cabral ex Borhidi* leaf

| Peak# | R.Time | Area | Area% | Height | Height% | A/H | Name |
|---|---|---|---|---|---|---|---|
| 1 | 3.641 | 253972 | 1.63 | 22584 | 0.79 | 11.25 | Oxetane, 2,4-dimethyl-, trans- |
| 2 | 4.966 | 1414451 | 9.09 | 318161 | 11.12 | 4.45 | 1-Butanol, 3-methyl- |
| 3 | 33.649 | 59581 | 0.38 | 34292 | 1.20 | 1.74 | 7-Azabicyclo[4.1.0]heptane |
| 4 | 36.499 | 1211535 | 7.78 | 158951 | 5.55 | 7.62 | n-Hexadecanoic acid |
| 5 | 36.704 | 86930 | 0.56 | 39642 | 1.38 | 2.19 | Methyl 2,4-dimethyltetradecanoate |
| 6 | 39.054 | 3841156 | 24.68 | 704109 | 24.60 | 5.46 | Phytol |
| 7 | 39.641 | 457499 | 2.94 | 64896 | 2.27 | 7.05 | Decane, 4,5-dibromo-, (R*,R*)- |
| 8 | 45.125 | 2892152 | 18.58 | 489789 | 17.11 | 5.90 | Phenol, 3-pentadecyl- |
| 9 | 45.174 | 492998 | 3.17 | 220043 | 7.69 | 2.24 | Androstan-3-ol, 9-methyl-, acetate, (3.beta.,5.al |
| 10 | 45.373 | 262824 | 1.69 | 69455 | 2.43 | 3.78 | 4-[4-[(2-Hydroxybenzoyl)amino]anilino]-4-oxo |
| 11 | 50.222 | 4591353 | 29.50 | 740513 | 25.87 | 6.20 | Squalene |
|  |  | 15564451 | 100.00 | 2862435 | 100.00 |  |  |

Library

*The Table showed the components of the extracts with squalene, phytol, and phenol, 3-pentadecyl- as the major components.

Table S3.2: Major Phytochemical composition of the ethanol extract of *Diodella sarmentosa (Sw) Bacigalupo El Cabral ex Borhidi* leaf and their Biological Activities

| S/N | R.TIME | PEAK AREA (%) | NAMES | BIOLOGICAL ACTIVITIES | REFERENCES |
|---|---|---|---|---|---|
| 1 | 4.966 | 9.09 | 1-Butanol,3-methyl- | Antimicrobial activities | Farina *et al.,* (2014). |
| 2 | 36.499 | 7.78 | n-Hexadecanoic acid | Anti-inflammatory, nematicide, pesticide, hemolytic, 5-Alpha reductase inhibitor, antifungal activities, and antibacterial activities | Bhawanker et al. (2013). |
| 3 | 39.054 | 24.68 | Phytol | Antimicrobial Activities; antioxidant, anticancer and antinociceptive activities. | Khasawneh *et al* (2011); Ammal and Bai (2013). |
| 4 | 45.125 | 18.58 | Phenol,3-pentadecyl- | Antioxidant | Correia *et al.,* (2006). |
| 5 | 50.222 | 29.50 | Squalene | Antioxidant, Antitumor | Kalvodona (2010). |

*The Table showed the biological activities of the major components of the extract.

| Strains | Dehydrogenase Activities (mg formazan/mg cell dry weight) |
|---|---|
| *Bacillus subtilis* | 1.064±0.14 |
| *Candidas spp* | 2.613 ± 0.10 |
| *Panicillium spp* | 2.869 ± 0.11 |

Table S3.3: Control Dehydrogenase Activities the of the Microbial Isolates

| Strains | $IC_{50}$ (µg/ml) | $R^2$ |
|---|---|---|
| *Bacillus subtilis* | 411 | 0.981 |
| *Candidas spp* | 275 | 0.954 |
| *Penicillium spp* | 322 | 0.948 |

Table S3.4: Threshold Inhibitory Concentration of Extract.

* *Penicillium spp* recorded the highest DHG activity.

*At $IC_{50}$ 275µg/ml, *Candidas spp* is the most sensitive to the extract.

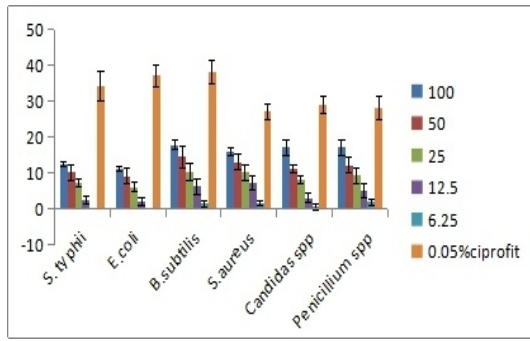

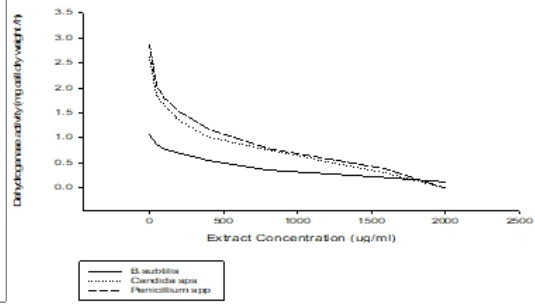

Figure S3.1: Diameter of Zone of Inhibition (mm) concentration (µg/ml).

Fig. S3.2: DHG activity of the organisms vs extract

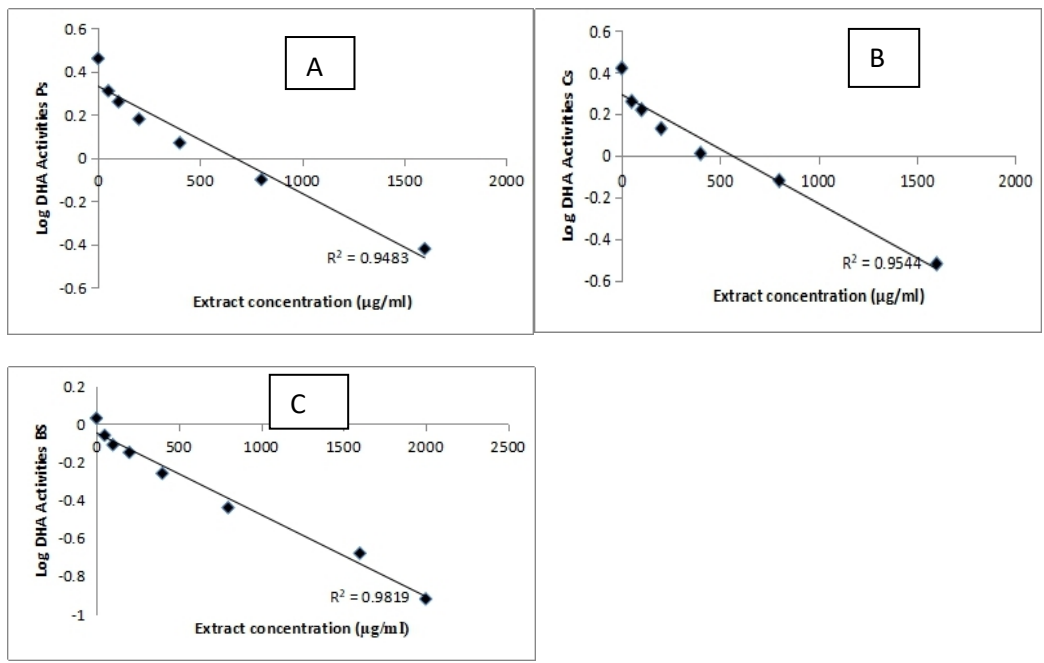

Figure S3.3A,B&C: Linearized plot of Log DHA Activities vs Extract Concentration (µg/ml)